\documentclass[11pt,a4paper,twoside]{article}
\usepackage{epsfig}
\usepackage[latin1]{inputenc}
\usepackage{graphicx,bbm}
\usepackage{amsmath,amsfonts,amssymb}
\usepackage{fleqn} 
\usepackage{mathrsfs}
\usepackage[footnotesize]{caption}

\begin{document}

\begin{titlepage}

\vspace*{-15mm}
\begin{flushright}
MPP-2008-37\\
\end{flushright}
\vspace*{0.7cm}

\begin{center}
{
\bf\LARGE Signals of CPT Violation and Non-Locality\\[2mm] 
in Future Neutrino Oscillation Experiments
}
\\[8mm]
S.~Antusch$^{\star}$
\footnote{E-mail: \texttt{antusch@mppmu.mpg.de}}, 
E.~Fernandez-Martinez$^{\star}$
\footnote{E-mail: \texttt{enfmarti@mppmu.mpg.de}},
\\[1mm]

\end{center}
\vspace*{0.50cm}
\centerline{$^{\star}$ \it 
Max-Planck-Institut f\"ur Physik (Werner-Heisenberg-Institut),}
\centerline{\it 
F\"ohringer Ring 6, D-80805 M\"unchen, Germany}
\vspace*{1.20cm}

\begin{abstract}

\noindent 
We investigate the sensitivities of future neutrino oscillation experiments for 
measuring the neutrino mass squared differences and leptonic mixing angles
independently with neutrinos and anti-neutrinos. We update the expected sensitivities
of Neutrino Factories to the ``atmospheric'' (anti-)neutrino parameters 
using an optimized setup. A dedicated $\beta$-Beam facility, 
in combination with a SPMIN reactor experiment, could give excellent sensitivities 
also to the ``solar'' parameters, for neutrinos and anti-neutrinos respectively.
A signal of a different mass matrix for neutrinos and anti-neutrinos
would imply CPT violation and non-locality of the underlying particle theory. 

\end{abstract}

\end{titlepage}

\newpage
\setcounter{footnote}{0}

\section{Introduction}
With the expected high sensitivities of envisioned neutrino facilities like Neutrino Factories 
\cite{Geer:1997iz}-\cite{Apollonio:2002en} or $\beta$-Beams \cite{Zucchelli:sa}, neutrino physics could enter a new era of precision \cite{Bandyopadhyay:2007kx}. One main goal of these experiments is the measurement of the yet unknown low energy parameters of the lepton sector accessible to oscillation physics, i.e.\ the Dirac CP phase $\delta$, the mixing angle $\theta_{13}$ and the sign of the ``atmospheric'' mass squared difference (corresponding to whether the mass spectrum is normally ordered or inverted). However, in addition to the determination of these parameters, future precision experiments may also find interesting signals of physics beyond the SM or even signals of violation of fundamental principles such as CPT invariance, Lorentz invariance or locality \cite{Kostelecky:2003cr}.  
One interesting signal of this type would be a difference between the masses and mixing angles measured by   neutrino oscillation experiments operating with neutrinos and anti-neutrinos. 

The CPT theorem \cite{cpttheorem} states that any local Quantum Field Theory (QFT) which is Lorentz invariant and has a Hermitean Hamiltonian must have CPT symmetry. From the derivation of the CPT theorem it also follows that CPT violation implies violation of Lorentz invariance. Bounds on CPT and Lorentz invariance violation from processes involving quarks and charged leptons are quite strong, for example from the $K^0 - \bar K^0$ system \cite{Yao:2006px}. On the other hand, CPT and Lorentz invariance are tested to a much less precision in the neutrino sector by the current experiments, as we will review below in section \ref{Sec:Present}. In addition to the less tight constraints, another motivation to look for signals of CPT and Lorentz invariance violation in the neutrino sector is the potentially different mechanism for generating the small neutrino masses compared to the masses of quarks and charged leptons, which might be especially sensitive to new physics.

One direct consequence of CPT invariance is that neutrinos and anti-neutrinos have the same masses and mixing angles. It has been pointed out in Ref.~\cite{Greenberg:2002uu} that a signal for the violation of this prediction, i.e.\ different masses and mixing angles for neutrinos and anti-neutrinos, would allow to draw the additional conclusion that in addition to CPT and Lorentz invariance also locality must be violated. More precisely, it has been shown that a difference between the masses for particles and anti-particles implies non-locality in the sense that field (anti-)commutators do not vanish for space-like distances and furthermore that the fields enter terms in the Lagrangian at different space-time points \cite{Greenberg:2002uu}. Non-locality is in general predicted by extensions of the SM towards a unified theory with gravity. Regarding CPT violation, it has been argued in Ref.~\cite{Kostelecky:1991ak} that non-local interactions in string theory might generate CPT violation close to present bounds. Specific scenarios of CPT breaking can also be realised in noncommutative geometries \cite{AmelinoCamelia:1999pm}. 
We would like to note, however, that if a signal of CPT violation via different masses of neutrinos and anti-neutrinos would be observed, it could be challenging to distinguish the possible intrinsic CPT violation and non-locality from ``fake'' signals, caused e.g.\ by non-standard matter effects. 
In any case, if the experimental data would point to different masses and/or mixing angles for neutrinos and anti-neutrinos, it would be an intriguing signal of physics beyond the SM.

Various aspects regarding CPT violation in the neutrino sector have been analysed in previous studies. For example, Ref.\ \cite{Bahcall:2002ia} has discussed the potential to test CPT by combining solar neutrino data and data from the KamLAND reactor experiment. The sensitivity of experiments with atmospheric neutrinos has been discussed in \cite{Datta:2003dg} and the prospects of the MINOS experiment has been studied in \cite{Rebel:2008th}. For the ``atmospheric'' parameters, high sensitivity to CPT violation could be achieved in Neutrino Factories \cite{Barger:2000iv,Bilenky:2001ka}. Additional tests might be possible with the data from supernova neutrinos \cite{Minakata:2005jy} and with neutrinoless double beta decay \cite{Barenboim:2002hx}. 
One motivation for the study of models of CPT violation for neutrino physics has been the LSND anomaly, where CPT violation in the neutrino sector was proposed as a solution \cite{Murayama:2000hm}. The case studied by most authors in this context is that of a CPT violating background which leads to CPT and Lorentz invariance violating local effective operators at low energy \cite{Colladay:1996iz}. Recently, a parameterisation for a three family oscillation analysis in the presence of specific types of CPT violation has been studied in \cite{Dighe:2008bu}. 

In order to search for a signal of CPT violation and non-locality in neutrino oscillation experiments a precise determination of the neutrino mass splittings and/or leptonic mixing angles, independently with neutrinos and anti-neutrinos, is desirable.  
Recently, significant progress has been made in the design of possible future neutrino oscillation facilities, for example regarding Neutrino Factories \cite{Bandyopadhyay:2007kx}. Furthermore, $\beta$-Beam facilities have been proposed which, in some aspects, could even allow more precise measurements than a Neutrino Factory. In addition, new reactor experiments (for $\bar\theta_{13}$ \cite{Ardellier:2006mn} as well as for precision measurements of $\bar\theta_{12}$ \cite{Bandyopadhyay:2003du}-\cite{Bandyopadhyay:2004cp}) have been envisioned.  
In this letter, we therefore investigate the combined sensitivity of such future neutrino oscillation experiments for measuring the ``solar'' and ``atmospheric'' mass squared differences and leptonic mixing angles independently with neutrinos and anti-neutrinos.
We update the expected sensitivities for the ``atmospheric'' mass squared difference and mixing angle with Neutrino Factories using an optimized setup and analyse how a dedicated $\beta$-Beam facility, in combination with a SPMIN reactor experiment \cite{Bandyopadhyay:2003du}-\cite{Bandyopadhyay:2004cp}, could improve the sensitivities for the ``solar'' mass splitting and mixing angle.

\section{Potential signals of CPT violation and non-locality}

In the following we denote parameters for anti-neutrinos with bars and parameters for neutrinos without bars. Using standard PDG parameterisation of the PMNS matrix \cite{Yao:2006px}, CPT invariance implies $m_i = \bar m_i$ and $\theta_{ij}=\bar \theta_{ij}$. 
As discussed above, a violation of these equalities, i.e.\ $m_i \not= \bar m_i$ for one neutrino mass eigenvalue or $\theta_{ij}\not=\bar \theta_{ij}$ for one of the mixing angles, if not confused with a ``fake'' signal of different new physics, would signal violation of CPT invariance (and Lorentz invariance) as well as non-locality of the underlying particle theory. 
To analyse the expected sensitivities to such signals of new physics beyond the SM, we study how well neutrino oscillation experiments can determine the neutrino masses and leptonic mixing angles for neutrinos and anti-neutrinos, i.e.\ using only data from oscillation of neutrinos and anti-neutrinos, respectively. In our analysis we treat the parameters $\theta_{ij},\bar \theta_{ij}$ and 
$m_i,\bar m_i$ as independent quantities and assume the standard three family oscillation formulae for neutrinos $\nu_\alpha$ and anti-neutrinos $\bar\nu_\alpha$, thereby testing the consistency of the CPT symmetric description.    
Bounds on CPT violation are presented 
as the constraints on differences $|\sin^2 \theta_{ij} - \sin^2 \bar{\theta}_{ij}|$ for the mixing angles and $|\Delta m^2_{ij} - \Delta \bar{m}^2_{ij}|$ for the mass squared differences (defined as $\Delta m^2_{ij} \equiv m_j^2 - m_i^2$ and $\Delta  \bar m^2_{ij} \equiv \bar m_j^2 -\bar m_i^2$) which the considered  experiments can impose.     
Before we turn to the discussion of the expected sensitivities of future neutrino oscillation facilities in this respect, we review the bounds on the difference between neutrino and anti-neutrino parameters from the present data. A graphical summary of the present bounds can be found in Fig.~55 of \cite{GonzalezGarcia:2007ib}.

\subsection{Present bounds (anti-)neutrino parameters}\label{Sec:Present}

Regarding the ``solar'' neutrino parameters $\theta_{12}$, $\bar\theta_{12}$ and $\Delta m^2_{21}$, $\Delta \bar{m}^2_{21}$, the most relevant data stems from experiments with solar neutrinos \cite{Cleveland:1998nv}-\cite{Ahmed:2003kj} and from the KamLAND experiment \cite{Eguchi:2002dm} observing anti-neutrinos from nuclear reactors. If CPT invariance is assumed, the complementarity between these two data sets allows for a quite good determination of the ``solar'' mixing angle as well as of the ``solar'' mass squared difference. However, separating into neutrino and anti-neutrino parameters shows that the bounds on their difference are comparatively weak.  
While solar neutrino data allows for a quite precise measurement of $\sin^2 \theta_{12}$, 
there is very little information on $\Delta m^2_{21}$. The reason for the good sensitivity on $\sin^2 \theta_{12}$ is that solar matter effects adiabatically convert the $\nu_e$ produced in the solar core into $\nu_2$ when leaving the sun, and thus detectors on earth can extract $|U_{e2}|^2=\sin^2 \theta_{12}$ by comparing the measured $\nu_e$ flux to the theoretically expected flux without oscillations or to the total neutrino flux measured through neutral current interactions. The sensitivity of the solar neutrino experiments to the mass splitting $\Delta m^2_{21}$ is very poor because the distance between earth and sun is so large that the oscillatory behaviour of the transition probability is averaged out. 
KamLAND, on the other hand, has a much shorter average baseline of $\sim 100$ Km. The (almost) vacuum oscillation signal of the reactor $\bar \nu_e$ and the measurement of the distortion of the anti-neutrino energy spectrum allows a precise measurement of $\Delta \bar m^2_{21}$. However, the overall normalization of the neutrino flux is rather difficult due to the fact that anti-neutrinos from many reactors (as well as from other sources) are detected by KamLAND. Consequently the constraints on $\sin^2 \bar\theta_{12}$ from KamLAND is comparatively weak. Furthermore, the near vacuum oscillations observed by KamLAND depend on the quantity $\sin^2 2 \bar\theta_{12}$, which implies that the octant of $\bar{\theta}_{12}$ cannot be determined from the present data on anti-neutrinos.
In summary, assuming that $\bar{\theta}_{12}$ lies in the first octant, the present bounds on CPT violation in the ``solar'' sector are at $3\sigma$
\begin{eqnarray}
|\sin^2 \theta_{12} - \sin^2 \bar{\theta}_{12}| &<& 0.3 \;,\\
|\Delta m^2_{21} - \Delta \bar{m}^2_{21}| &<& 1.1 \times 10^{-4} \:\mbox{eV}^2\;.
\label{present_solar}
\end{eqnarray}

Regarding the ``atmospheric'' parameters $\theta_{23}$, $\bar\theta_{23}$ and $\Delta m^2_{31}$, $\Delta \bar{m}^2_{31}$, the most important experimental data stems from atmospheric neutrinos observed by SuperKamiokande ~\cite{Fukuda:1998mi,Ambrosio:2001je} and from the accelerator experiments sensitive to the ``atmospheric'' sector, K2K \cite{Aliu:2004sq} and MINOS \cite{Michael:2006rx}. K2K and MINOS operate with neutrinos and the atmospheric signal is dominated by the neutrino event rates due to the larger cross section. Therefore, the measurements of $\sin^2 \bar{\theta}_{23} $ and in particular of $\Delta \bar{m}^2_{31}$ are much less precise than the ones of $\sin^2 \theta_{23}$ and  $\Delta m^2_{31}$ (see Fig.~55 of \cite{GonzalezGarcia:2007ib}). The present bounds on CPT violation for the ``atmospheric'' sector are at $3\sigma$~\cite{Bandyopadhyay:2007kx,GonzalezGarcia:2004wg}:
\begin{eqnarray}
|\sin^2 \theta_{23} - \sin^2 \bar{\theta}_{23}| &<& 0.45 \;,\\
|\Delta m^2_{31} - \Delta \bar{m}^2_{31}| &<& 1 \times 10^{-2} \:\mbox{eV}^2\;.
\label{present_atmo}
\end{eqnarray}

Finally the bounds on the unknown 1-3 mixing angles $\theta_{13}$ and $\bar\theta_{13}$ for neutrinos and anti-neutrinos, respectively, are similar. The bound $ \sin^2 \bar{\theta}_{13} < 0.1$ (at $3 \sigma$) is obtained from the CHOOZ experiment \cite{Apollonio:2002gd} and a bound of $ \sin^2 \theta_{13} < 0.3$ can be derived from the combined data of experiments on solar, atmospheric and accelerator neutrino oscillations \cite{GonzalezGarcia:2007ib}. From the present data, without a measurement of neither $\theta_{13}$ nor $\bar\theta_{13}$, the bound on the difference (at $3 \sigma$) is given by
\begin{eqnarray}
|\sin^2 \theta_{13} - \sin^2 \bar{\theta}_{13}| &<& 0.3 \;.
\label{present_t13}
\end{eqnarray}
We will now discuss how these bounds (or equivalently the sensitivities for discovering a signal) might be improved in future neutrino oscillation facilities.

\subsection{Strategies for improving the sensitivities}

As has been discussed in previous studies \cite{Bilenky:2001ka}, the bounds on $|\sin^2 \theta_{23} - \sin^2  \bar{\theta}_{23}| $ and $|\Delta m^2_{31} - \Delta \bar{m}^2_{31}|$ can be strongly improved at a Neutrino Factory facility \cite{Geer:1997iz}-\cite{Apollonio:2002en}. Excellent sensitivities to $|\sin^2 \theta_{13} - \sin^2  \bar{\theta}_{13}| $ are also to be expected since the determination of $\theta_{13}$ is one of the main goals of the Neutrino Factory. In the context of the IDS, {\em International Design Study of the Neutrino Factory}, a baseline for the Neutrino Factory acceleration complex and detection systems has recently been defined \cite{Bandyopadhyay:2007kx,IDS}. This baseline setup would store 25 GeV muons, whose (anti-)neutrino fluxes aim at two 50 Kton Magnetized Iron Neutrino Detectors (MIND)
located at $L \sim 4000$ Km and $L \sim 7500$ Km from the source, respectively. The goal luminosity for such a facility is $5 \times 10^{20}$ useful muon decays per year per polarity per baseline. The efficiencies, backgrounds and energy resolution of the MIND detector when exposed to such a beam have been studied in \cite{Abe:2007bi}. We will present the up-dated bounds with these specifications in section \ref{Sec:Future}. In comparison, little work has been done so far with respect to the improvement of the bounds on the ``solar'' parameters, and we will therefore focus on this issue in the remainder of this section.  

As discussed in section \ref{Sec:Present}, the present uncertainty on $|\sin^2  \theta_{12} - \sin^2  \bar{\theta}_{12}| $ is dominated by the comparatively low precision on $ \sin^2 \bar{\theta}_{12} $, while the error on $|\Delta m^2_{21} - \Delta \bar{m}^2_{21}|$ stems from the weak constraints on $\Delta m^2_{21}$. Improving the former is much easier than the latter. A dedicated reactor experiment placed at the minimum of the $\bar{\nu}_e$ survival probability (SPMIN) could provide an excellent measurement of $\sin^2 \bar{\theta}_{12}$ \cite{Bandyopadhyay:2003du}-\cite{Bandyopadhyay:2004cp}. A precise determination of 
$\Delta m^2_{21}$, however, is much more challenging. The observation of the oscillations driven by the small ``solar'' splitting requires large values of $L/E$. The neutrino charged current (CC) interaction, however, is very suppressed at low energies since the neutrons with which they can interact are confined in nuclei and a minimum threshold energy is normally required for the interaction to occur. 
Moreover, while nuclear reactors provide a convenient source of electron anti-neutrinos, it is challenging to produce low energy electron neutrinos in large amounts on earth.

\begin{figure}
\vspace{-0.5cm}
\begin{center}
\hspace{-0.55cm} \epsfxsize7cm\epsffile{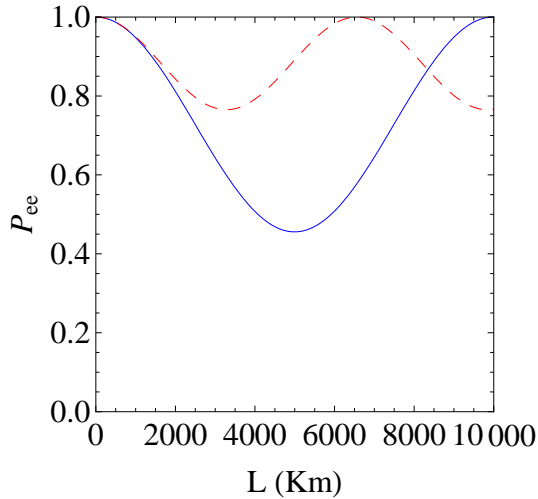} 
\caption{
$\nu_e$ (solid) and $\bar{\nu}_e$ (dashed) oscillation probability for $E=0.4$ GeV as a function of the baseline.
}
\label{prob}
\end{center}
\end{figure}

One possibility to improve the determination of $\Delta m^2_{21}$ would be an intense $\nu_e$ beam of $E_\nu \simeq 0.4$GeV from the $\beta$ decay of $^{18}$Ne ions accelerated to $\gamma = 100$ at a $\beta$-Beam facility \cite{Zucchelli:sa},\cite{Bouchez:2003fy}-\cite{Campagne:2006yx}. However, to optimize the determination of $\Delta m^2_{21}$ and $\theta_{12}$ with neutrinos, we consider also a very long baseline of $4000$ Km to a Mton class water Cerenkov detector \cite{Itow:2001ee}-\cite{Diwan:2006qf}. Somewhat less sensitivity can be reached with a more conventional 750 Km baseline. We have also studied the possible measurements achievable with the $\bar{\nu}_e$ that this facility could also provide from the decays of $^6$He. However we found that the performance of the anti-neutrino beam was significantly worse. This asymmetry is caused by the different matter effects for neutrinos and anti-neutrinos. In particular, the electron (anti-)neutrino survival probabilities for $\theta_{13}=0$ are given by (see e.g.~\cite{Akhmedov:2004ny}):
\begin{equation}
P_{ee} = 1- \frac{\sin^2 2 \theta_{12}}{s}\sin^2\left(\frac{\Delta m^2_{21}L}{4E}s\right)~,
\end{equation}
where 
\begin{equation}
s=\sqrt{\sin^2 2 \theta_{12} + (\cos 2 \theta_{12} \pm A)^2}
\label{matter}
\end{equation}
\noindent
with $A = \frac{V2E}{\Delta m^2_{21}}$ and where the $\pm$ signs apply to the probabilities of anti-neutrinos and neutrinos, respectively. Thus, matter effects reduce the amplitude of the oscillation probability and increase the oscillation frequency. Due to the $+$ sign in Eq. (\ref{matter}) for anti-neutrinos the effect is enhanced, the oscillation amplitude is smaller and a less precise reconstruction of the parameters follows. This situation is illustrated in Fig.~\ref{prob}. The solid and dashed lines correspond to the $\nu_e$ and $\bar{\nu}_e$ survival probabilities respectively. Consequently, present bounds on $\Delta \bar{m}^2_{21}$ and $\sin^2 \bar{\theta}_{12} $ will not be improved  by this setup. On the other hand, Eq.~(\ref{matter}) is no longer symmetric under the octant of $\bar{\theta}_{12}$. Changing the octant has the same effect as changing the sign of $A$ or, equivalently, interchanging the neutrino and anti-neutrino curves in Fig.~\ref{prob}. This means that the $\bar{\nu}_e$ from the $^6$He decay could distinguish the octant of $\bar{\theta}_{12}$ which would not be possible even at a dedicated SPMIN experiment for a precise measurement of $\sin^2 2\bar{\theta}_{12} $ due to the small matter effects. 

The $\beta$-Beam setup with 4000 Km baseline is thus optimized for the measurement of ``solar'' parameters through $\nu_e$ disappearance, however it is unrealistic to expect that such a dedicated facility would ever be built. As already mentioned above, we have therefore also considered a setup with a shorter, more conventional baseline of $750$ Km. This is the baseline usually chosen for the $\gamma = 350$ version of the $\beta$-Beam, which could provide excellent sensitivities to CP-violation \cite{BurguetCastell:2003vv} even outperforming the Neutrino Factory for some regions of the parameter space. A $\gamma = 100$ run of the $\beta$-Beam at this baseline, such as the one we consider here to improve the measurement on $\Delta m^2_{21}$, can improve
the sensitivity to the mass hierarchy and allow to solve degeneracies when combined with the higher $\gamma$ run \cite{Donini:2006dx}.

\subsection{Expected sensitivities of future facilities}\label{Sec:Future}

For the experimental setup we consider the ``IDS baseline'' Neutrino Factory \cite{IDS}, which provides excellent sensitivities to the ``atmospheric'' parameters and $\theta_{13}$ with both neutrino and anti-neutrino beams. In this facility intense (anti-)neutrino beams would be produced from the decay of $25$ GeV (anti-)muons with a luminosity of $5 \times 10^{20}$ muon decays per year per muon polarity. This beams illuminate two identical 50 Kton iron calorimeters located at 4000 Km and 7000 Km. 
We consider 5 years of data taking with each muon polarity. 
We did not consider the Opera-like detector to observe the silver channel \cite{Donini:2002rm} since it did not improve significantly any bound. 

In order to improve the present bounds on the ``solar'' sector we also considered a $\beta$-Beam facility producing electron (anti-)neutrinos from the decay of $^{18}$Ne ($^{6}$He) ions accelerated to $\gamma = 100$. We considered 5 years of data taking with luminosities of $10^{19}$ ion decays per year \cite{Lindroos2} for both ions. These neutrino beams would be detected at a Mton class water Cerenkov detector at a 750 or 4000 Km baseline. To describe the detector efficiencies and backgrounds when exposed to these beams we followed Ref.~\cite{Campagne:2006yx}.

Finally a reactor experiment with an exposure of $60$ GW$\cdot$Kton$\cdot$yr to a detector located at $60$ Km distance was considered to improve the measurement of the ``solar'' parameters with anti-neutrinos. We considered a 5\% systematic error in the normalization of the signal and the energy resolution of the detector was set to $\sigma = 0.05 \times E$.
The signal was distributed in 10 energy bins in the range between $0.0018$ and $0.008$ GeV.

\begin{table}
\begin{center}
\begin{tabular}{|c|c|c|c|c|}\hline
\vphantom{$\frac{f^2}{f^2}$} quantity & present bound & future ($\beta$B 4000 Km) & future ($\beta$B 750 Km)\\ \hline
\vphantom{$\frac{f^2}{f^2}$} $|\sin^2 \theta_{12} - \sin^2 \bar{\theta}_{12}|$ & 0.3 & 0.14 & 0.14 \\\hline  
\vphantom{$\frac{f^2}{f^2}$} $|\sin^2 \theta_{13} - \sin^2 \bar{\theta}_{13}|$ & 0.3 & $5.7 \times 10^{-4}$ & $5.7 \times 10^{-4}$ \\\hline  
\vphantom{$\frac{f^2}{f^2}$} $|\sin^2 \theta_{23} - \sin^2 \bar{\theta}_{23}|$ & 0.45 & 0.043 & 0.044 \\\hline  
\vphantom{$\frac{f^2}{f^2}$} $|\Delta m^2_{21} - \Delta \bar{m}^2_{21}| $ & $1.1\times 10^{-4} \:\mbox{eV}^2$ & $1.3\times 10^{-5} \:\mbox{eV}^2$ & $2.2\times 10^{-5} \:\mbox{eV}^2$  \\\hline 
\vphantom{$\frac{f^2}{f^2}$} $|\Delta m^2_{31} - \Delta \bar{m}^2_{31}| $ & $1 \times 10^{-2} \:\mbox{eV}^2$ & $2.6\times 10^{-5} \:\mbox{eV}^2$  & $3.3\times 10^{-5} \:\mbox{eV}^2$\\\hline 
\end{tabular}
\end{center}
\caption{Present bounds and expected future experimental sensitivities to differences between masses and mixing angles for neutrinos and anti-neutrinos at 3$\sigma$. The considered future facilities are explained in the main text. \label{tab:summary}}
\end{table}

For the numerical analysis we independently combined the neutrino and anti-neutrino data from the Neutrino Factory, SPMIN and $\beta$-Beam setups to derive measurements on the neutrino and anti-neutrino parameters. Since the combination of the three facilities allows to constrain all the oscillation parameters no prior information on any of them was assumed except for the measurement that will not be improved by this setup, namely $\theta_{12}$ from solar neutrino oscillations, for which we assumed a $1 \sigma$ uncertainty of 
10 \%. A 5 \% uncertainty in the PREM density profile was also assumed. The Globes 3.0 \cite{Globes} software was used to perform the numerical analysis. The following (CPT conserving) input values were assumed for the oscillation parameters:
$\sin^2 \theta_{12}=0.3$, 
$\sin^2 \theta_{23}=0.5$, 
$\sin^2 \theta_{13}=0$, 
$\Delta m^2_{21}=7.6 \times 10^{-5} \:\mbox{eV}^2$, 
$\Delta m^2_{31}=2.5 \times 10^{-3} \:\mbox{eV}^2$.

\begin{figure}
\vspace{-0.5cm}
\begin{center}
\begin{tabular}{cc} 
\hspace{-0.55cm} \epsfxsize6.5cm\epsffile{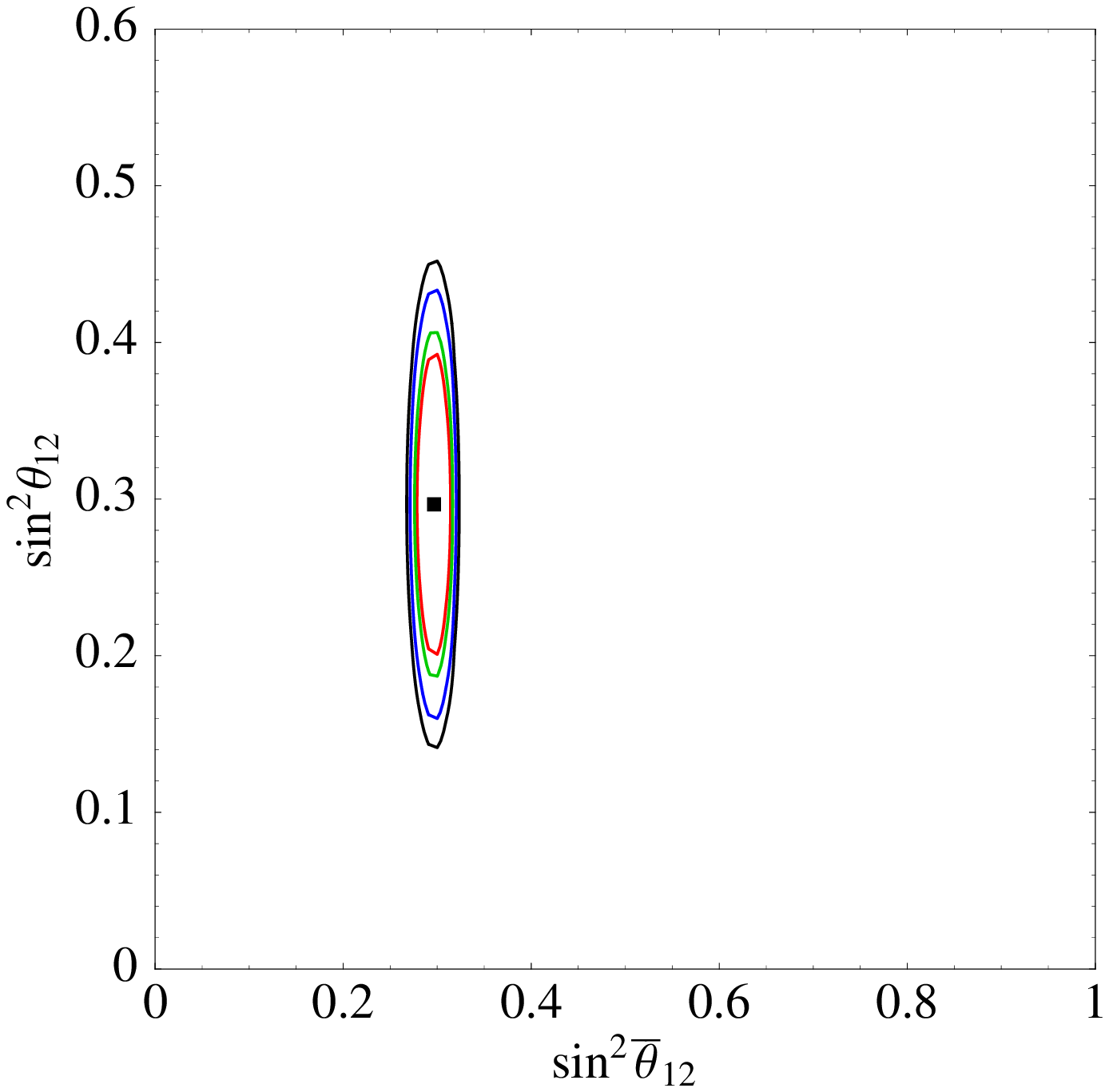} & 
                 \epsfxsize6.5cm\epsffile{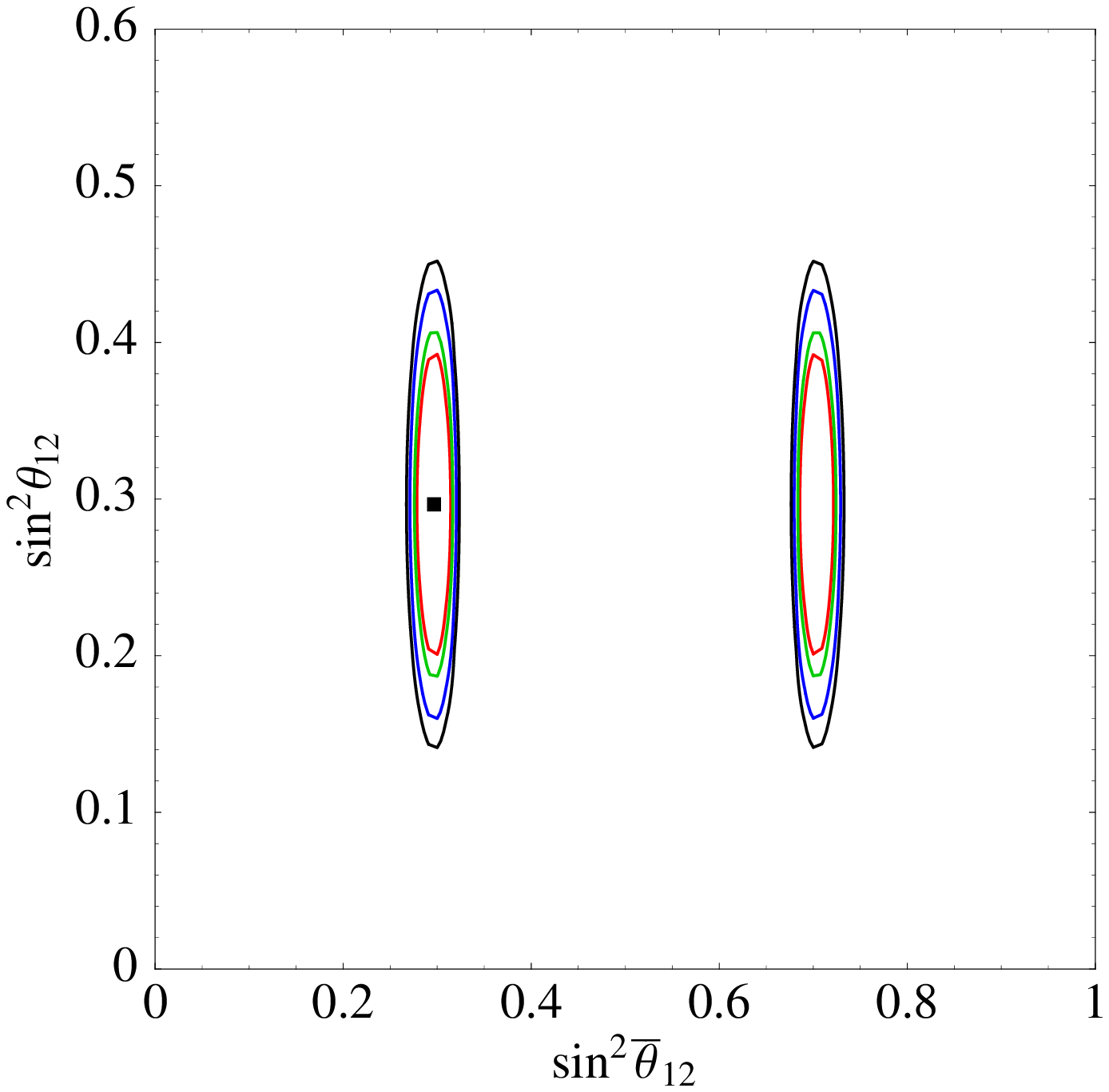} \\
\hspace{-0.55cm} \epsfxsize6.5cm\epsffile{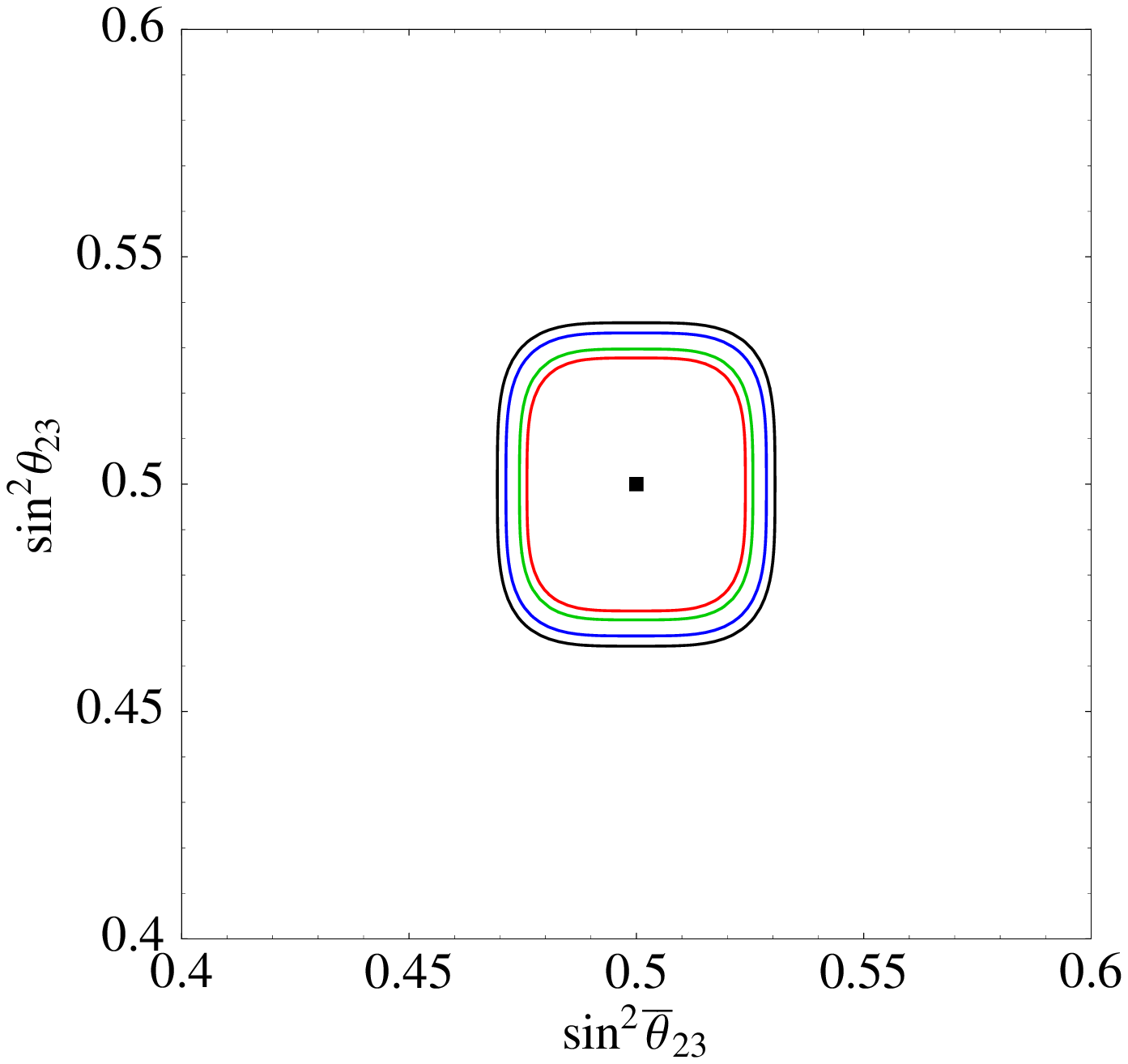} & 
                 \epsfxsize6.5cm\epsffile{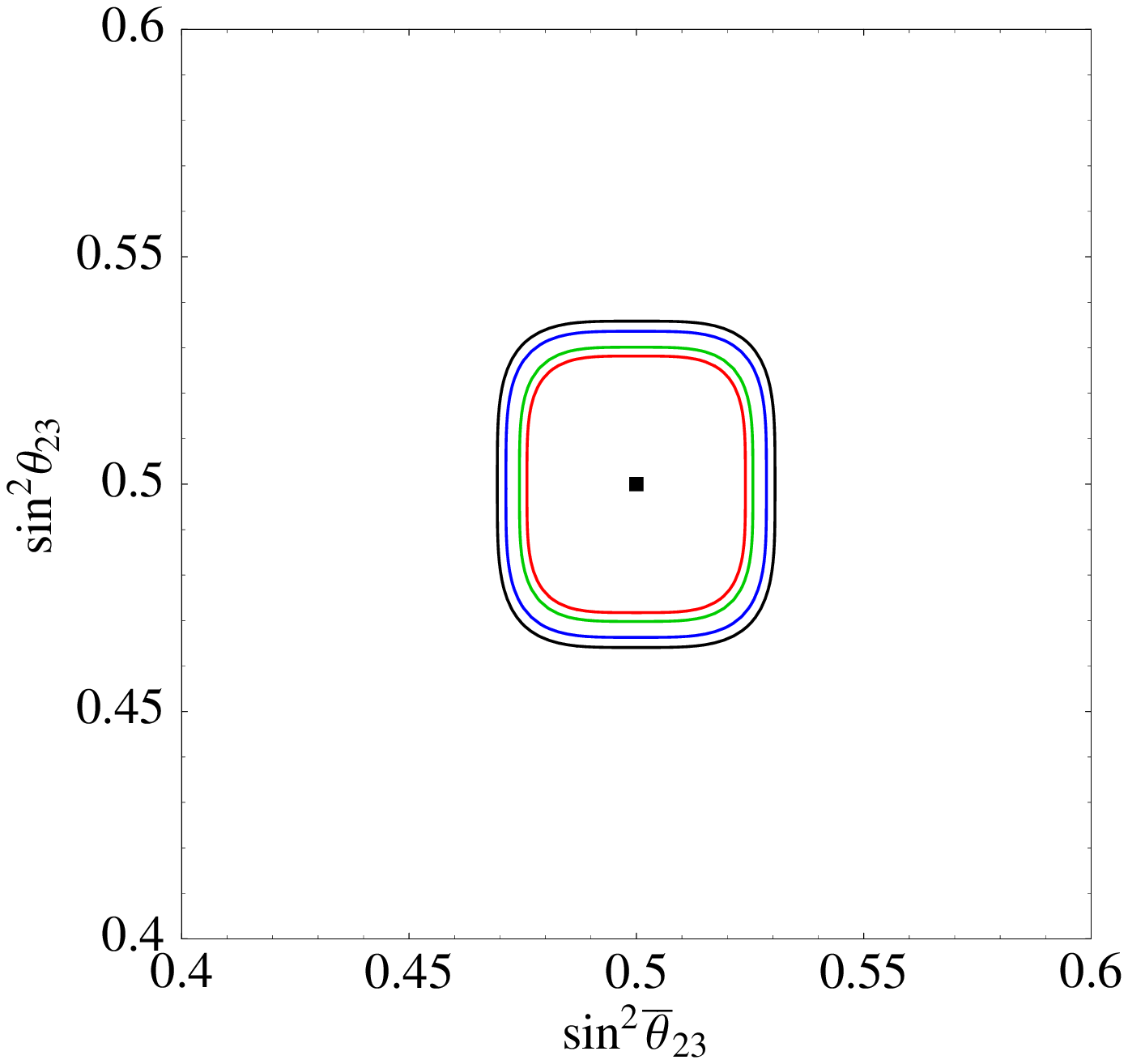} \\
\hspace{-0.55cm} \epsfxsize6.5cm\epsffile{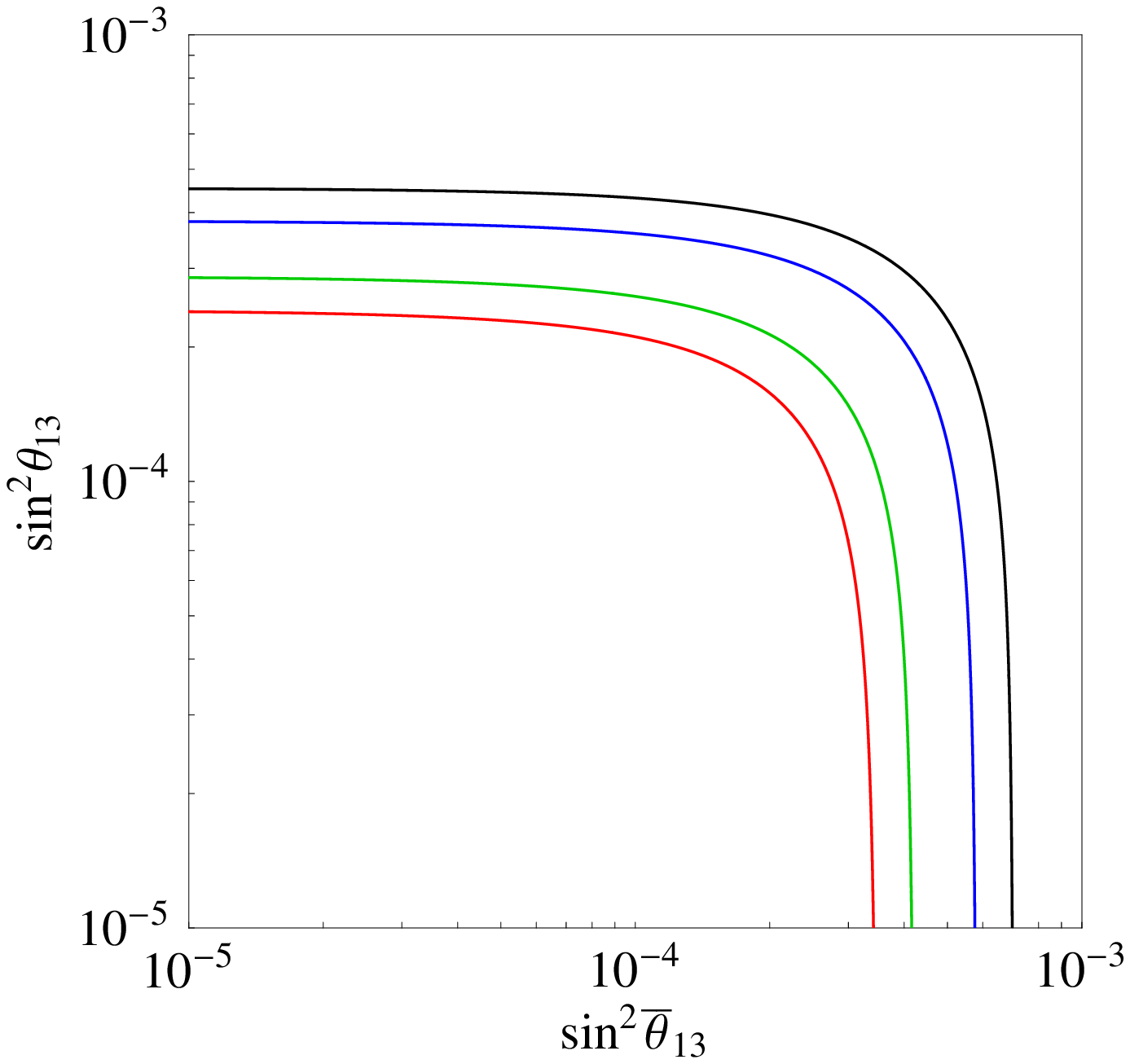} & 
                 \epsfxsize6.5cm\epsffile{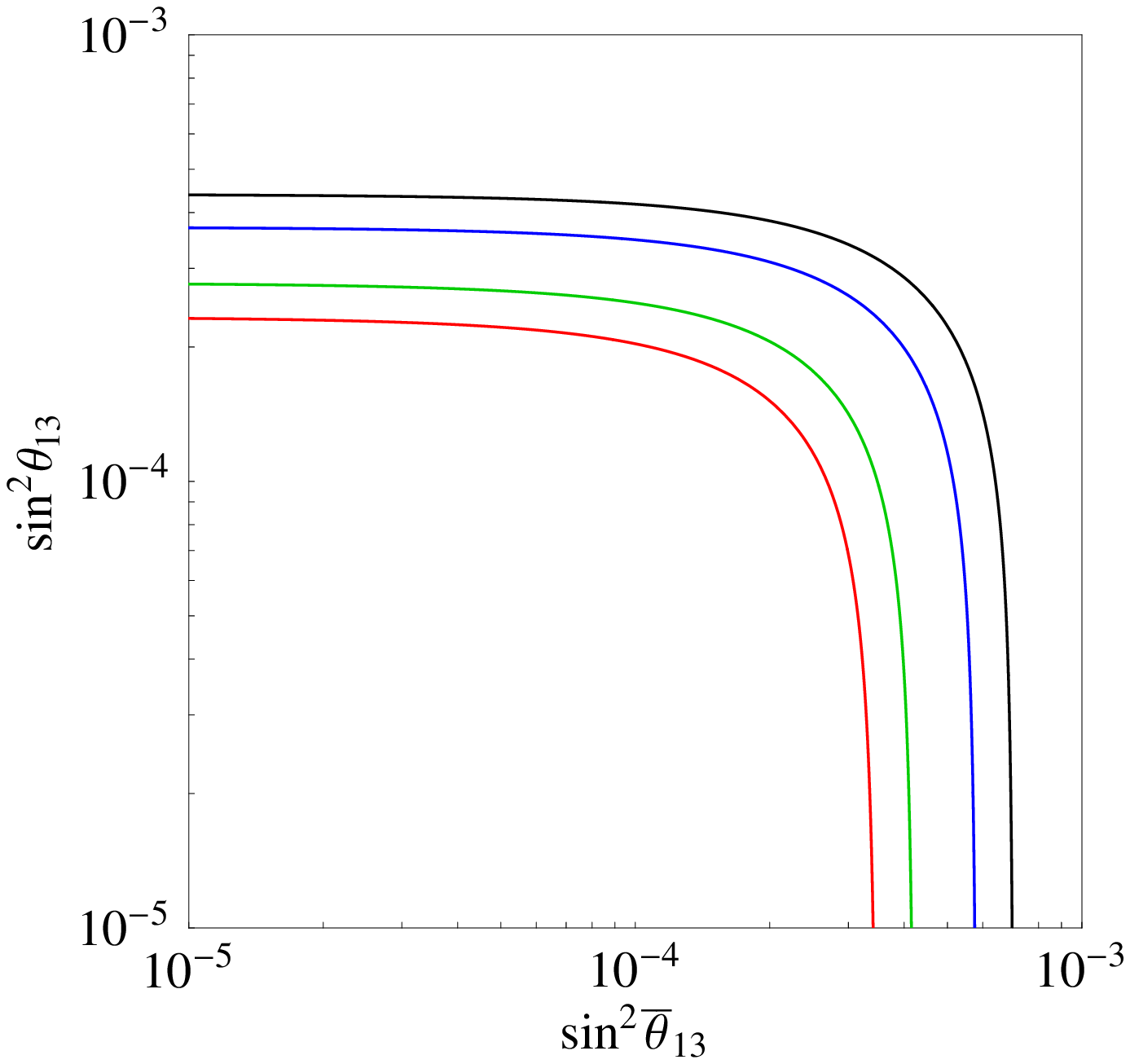} 
\end{tabular}
\caption{
$90 \%$, $95 \%$, $99 \%$ and $3 \sigma$ contours for the measurements of $\sin^2 \theta_{12}$,  $\sin^2 \theta_{23}$,  $\sin^2 \theta_{13}$,
with either neutrino data (vertical axes) or anti-neutrino data (horizontal axes) alone. The left (right) panels show the constraints achievable with the 4000 Km (750 Km) baseline for the $\beta$-Beam.}
\label{fig1}
\end{center}
\end{figure}

\begin{figure}
\vspace{-0.5cm}
\begin{center}
\begin{tabular}{cc} 
\hspace{-0.55cm} \epsfxsize6.5cm\epsffile{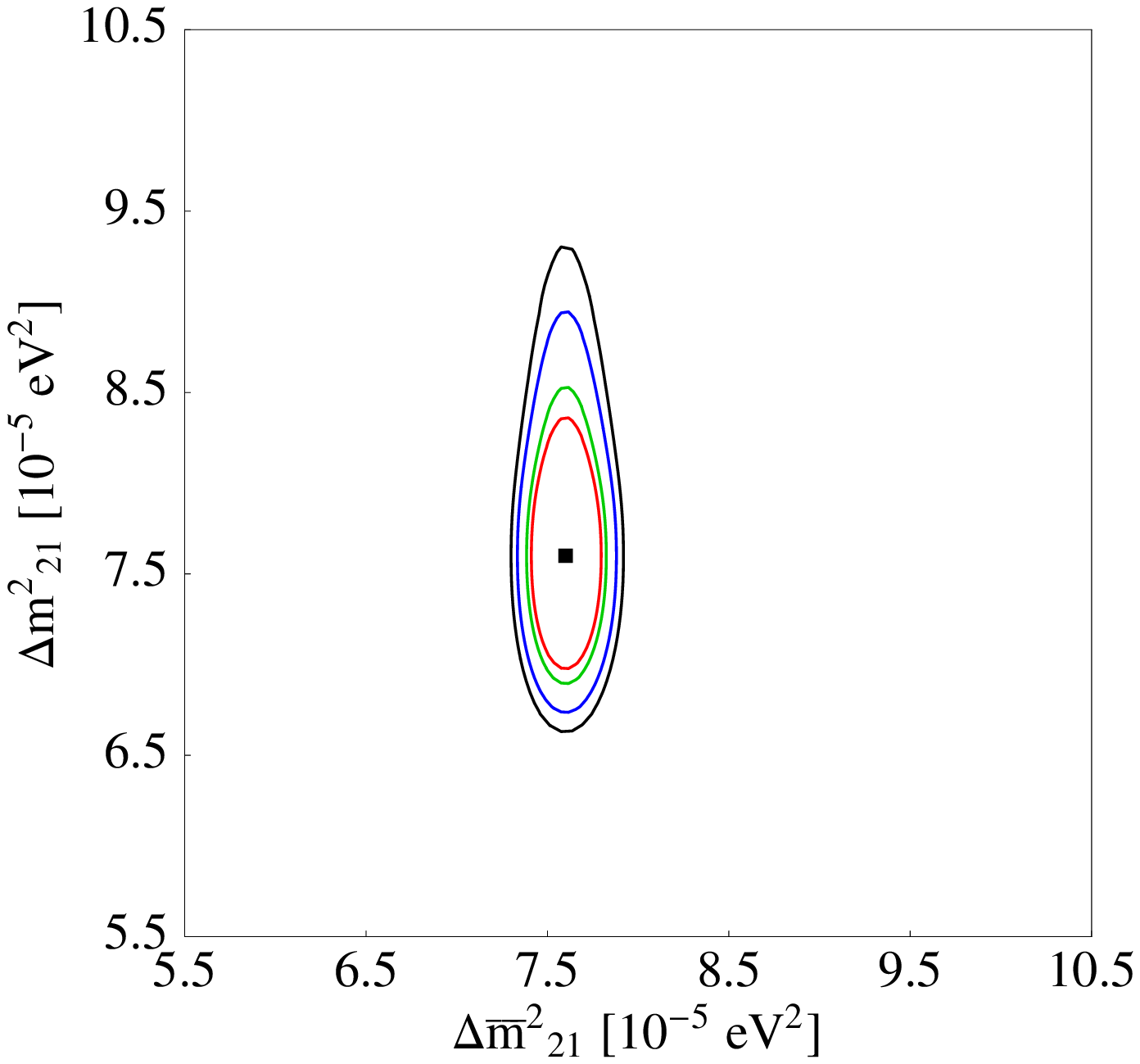} & 
                 \epsfxsize6.5cm\epsffile{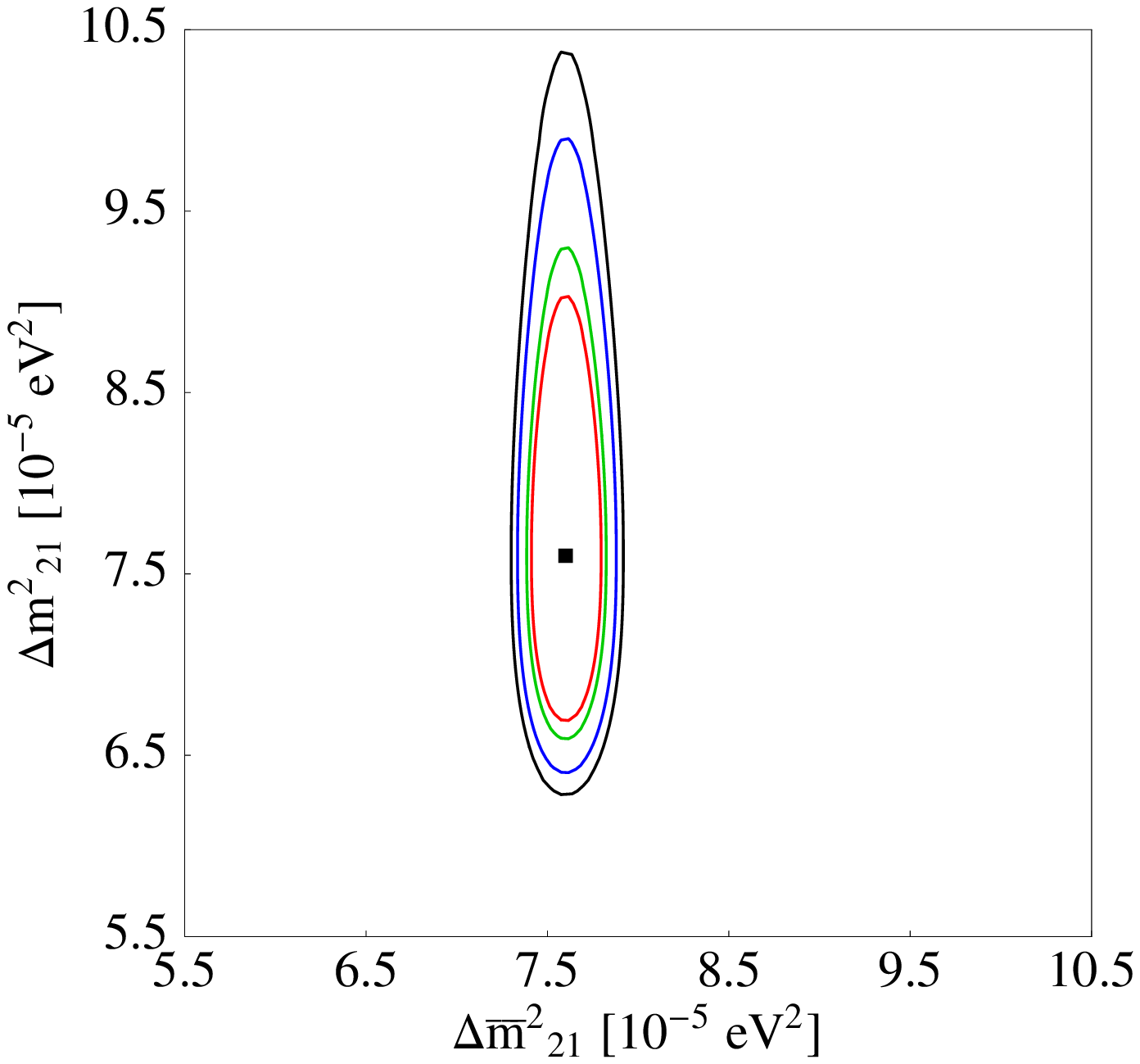} \\
\hspace{-0.55cm} \epsfxsize6.5cm\epsffile{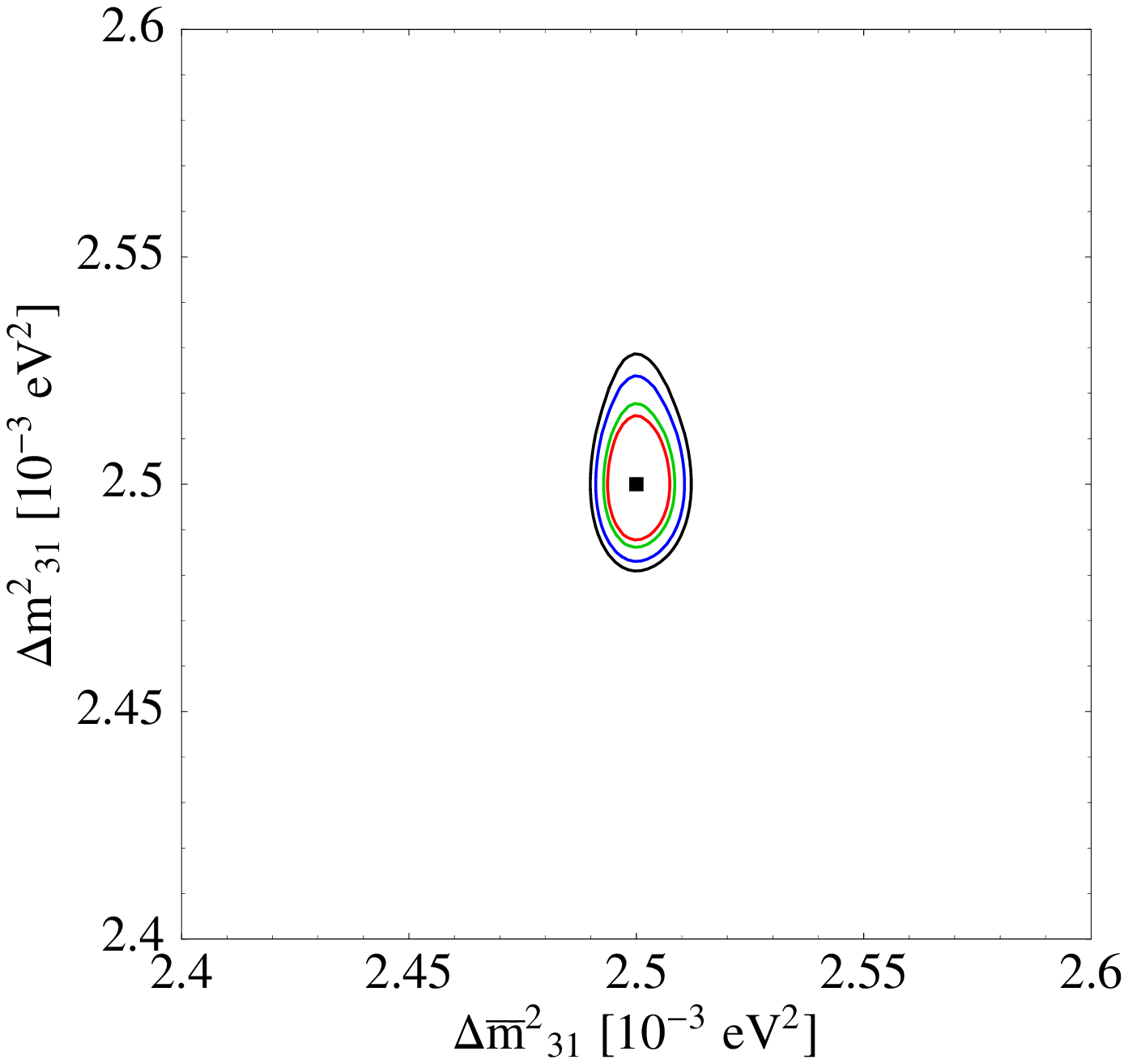} & 
                 \epsfxsize6.5cm\epsffile{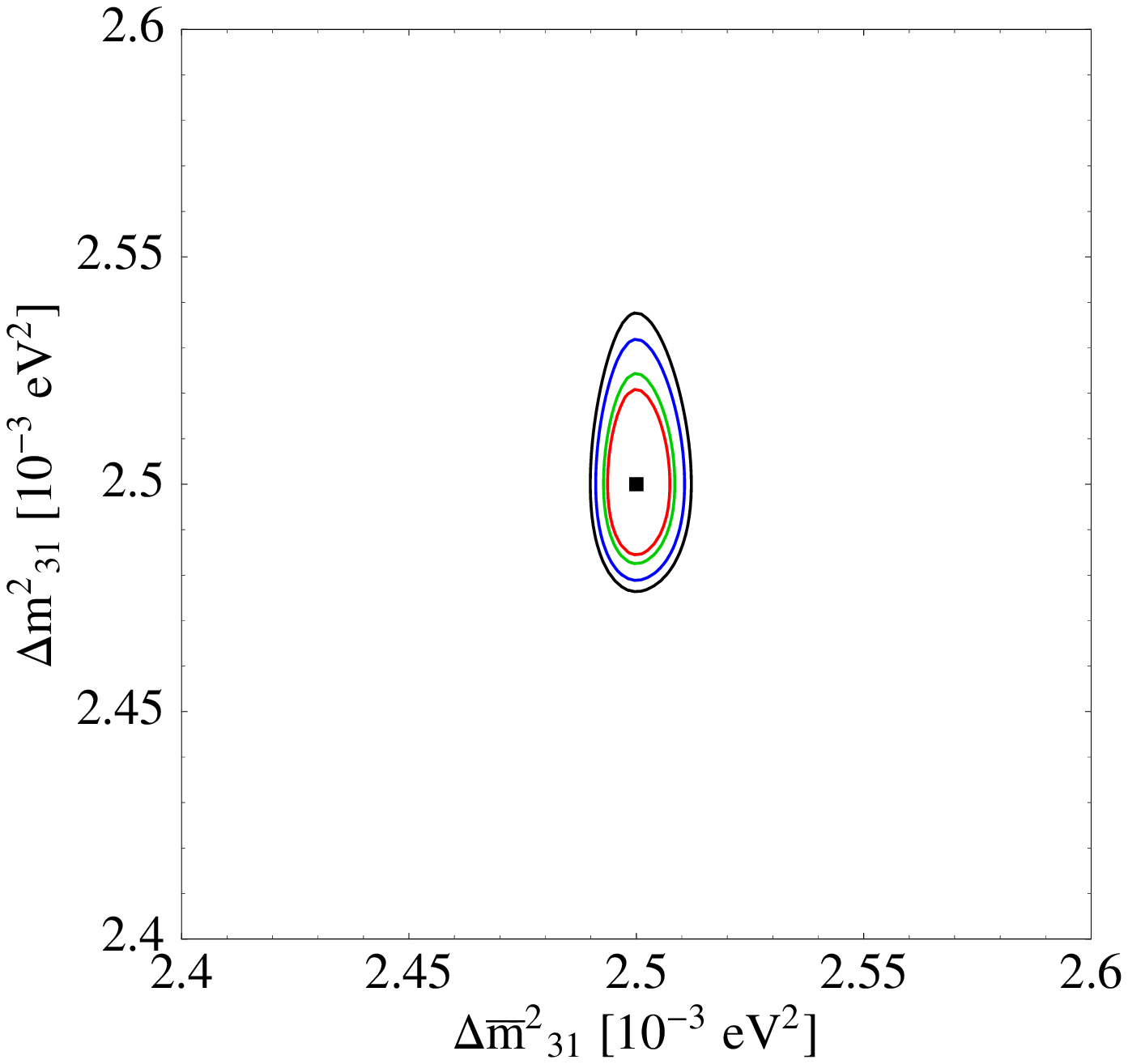} 
\end{tabular}
\caption{
$90 \%$, $95 \%$, $99 \%$ and $3 \sigma$ contours for the measurements of 
$\Delta m^2_{21}$ and $\Delta m^2_{31}$ with either neutrino data (vertical axes) or anti-neutrino data (horizontal axes) alone. The left (right) panels show the constraints achievable with the 4000 Km (750 Km) baseline for the $\beta$-Beam.}
\label{fig2}
\end{center}
\end{figure}

%
Figures \ref{fig1} and \ref{fig2} show the 90 \%, 95 \%, 99 \% and $3 \sigma$ contours for the constraints on $\sin^2 \theta_{12}$,  $\sin^2 \theta_{23}$, $\sin^2 \theta_{13}$, $\Delta m^2_{21}$ and $\Delta m^2_{31}$ with either neutrino data (vertical axes) or anti-neutrino data (horizontal axes) alone. The left (right) panels contain the constraints achievable with the 4000 Km (750 Km) baseline for the $\beta$-Beam. Comparing it with Fig.~55 of \cite{GonzalezGarcia:2007ib} the dramatic improvement on the constraints on the ``atmospheric'' parameters with the Neutrino Factory is manifest. The constraints for $\sin^2 \theta_{23}$ and $\sin^2 \bar{\theta}_{23}$ shrink by about one order of magnitude, while the improvement in $\Delta m^2_{31}$ and $\Delta \bar{m}^2_{31}$ is about two and three orders of magnitude respectively. The bounds on CPT violation that could be derived from this precision measurements including the 750 Km baseline $\beta$-Beam would be:
\begin{eqnarray}
|\sin^2 \theta_{23} - \sin^2 \bar{\theta}_{23}| &<& 0.043 \;,\\
|\Delta m^2_{31} - \Delta \bar{m}^2_{31}| &<& 3.3\times 10^{-5} \:\mbox{eV}^2\;.
\label{future_atmo}
\end{eqnarray}
With such a Neutrino Factory setup, the constraints on both $\sin^2 \theta_{13}$ and $\sin^2 \bar{\theta}_{13}$ could also be significantly improved by almost three orders of magnitude to
\begin{eqnarray}
|\sin^2 \theta_{13} - \sin^2 \bar{\theta}_{13}| &<& 5.7 \times 10^{-4}\;.
\label{future_t13}
\end{eqnarray}

In the ``solar'' sector the improvements are more modest, especially for the shorter $\beta$-Beam baseline.
For the 4000 Km baseline matter effects allow to measure the octant of $ \bar{\theta}_{23}$, but at 750 Km their strength is not sufficient for this task, as can be seen in the top panels of Fig.~\ref{fig1}. With neutrinos an improvement of $\Delta m^2_{21}$ by an order of magnitude could be accomplished with the $\beta$-Beam and some improvement in the lower bound of $\sin^2 \theta_{12}$ could also be achieved. These  improved measurements could test CPT invariance to the level of 
\begin{eqnarray}
|\sin^2 \theta_{12} - \sin^2 \bar{\theta}_{12}| &<& 0.14 \;,\\
|\Delta m^2_{21} - \Delta \bar{m}^2_{21}| &<& 1.3\times 10^{-5} \:\mbox{eV}^2\;,
\label{future_solar}
\end{eqnarray}
for the 4000 Km baseline $\beta$-Beam or
\begin{eqnarray}
|\sin^2 \theta_{12} - \sin^2 \bar{\theta}_{12}| &<& 0.14 \;,\\
|\Delta m^2_{21} - \Delta \bar{m}^2_{21}| &<& 2.2\times 10^{-5} \:\mbox{eV}^2\;,
\label{future_solar2}
\end{eqnarray}
for the 750 Km baseline $\beta$-Beam. In the latter case the octant of $\bar{\theta}_{12}$ is not measured. $\bar{\theta}_{12}$ has been assumed to lie in the first octant in order to compare with Eq.~(\ref{present_solar}). 
A summary of present and possible future bounds is presented in Tab.~\ref{tab:summary}. 

\section{Discussion and Conclusions}
In this study we have investigated the sensitivity of future neutrino oscillation experiments 
for measuring the ``solar'' and ``atmospheric'' mass squared differences and leptonic mixing angles independently with neutrinos and anti-neutrinos.
If a difference between the parameters for neutrinos and anti-neutrinos would be established, it would imply CPT (and Lorentz invariance) violation as well as non-locality. 

To improve the present bounds on this form of CPT violation we have considered three types of possible future neutrino oscillation experiments: an optimized Neutrino Factory, a $\gamma = 100$ $\beta$-Beam (with 750 Km or 4000 Km baseline) pointing to a Mton water Cerenkov detector and a SPMIN reactor experiment at the minimum of the $\bar{\nu}_e$ survival probability. The possible improvements with respect to the present bounds are summarised in Tab.~\ref{tab:summary}.

Regarding the ``atmospheric'' parameters and $\theta_{13}$ (vs.~$\bar\theta_{13}$), a dramatic improvement could be accomplished with a Neutrino Factory operating with neutrinos as well as with anti-neutrinos. The sensitivity on $|\sin^2 \theta_{13} - \sin^2 \bar{\theta}_{13}|$ could be improved almost three orders of magnitude, the sensitivity on $|\sin^2 \theta_{23} - \sin^2 \bar{\theta}_{23}|$ by one order of magnitude and the sensitivity on $|\Delta m^2_{31} - \Delta \bar{m}^2_{31}|$ by more than two orders of magnitude.

To improve the sensitivity with respect to the ``solar'' parameters, the combination of the $\beta$-Beam with the SPMIN reactor experiment could also be very successful. While the $\beta$-Beam experiment could improve the measurements of the parameters with neutrinos (mainly $\Delta m^2_{21}$) and determine the octant of $\bar{\theta}_{12}$, the SPMIN experiment could further improve the measurement of $\bar\theta_{12}$.   
While the gain in sensitivity on $|\sin^2 \theta_{12} - \sin^2 \bar{\theta}_{12}|$ amounts about 50 \%, the sensitivity on $|\Delta m^2_{21} - \Delta \bar{m}^2_{21}|$ could be improved by about one order of magnitude. 

In summary, future neutrino oscillation experiments have the potential to measure neutrino mass squared differences and leptonic mixing angles separately with neutrinos and anti-neutrinos to high precision and may detect possible differences between neutrino and anti-neutrino parameters. 
Such a difference would be a clear signal of new physics beyond the Standard Model of particle physics. If other types of new physics which may lead to ``fake'' signals, e.g.\ non-standard matter effects, could be excluded it would signal CPT (and Lorentz invariance) violation and also a non-local nature of the underlying particle theory.

\section*{Acknowledgements}
We would like to thank Andrea Donini for helpful discussions and Patrick Huber for kindly providing the description of the IDS baseline for the Neutrino Factory. This work was partially supported by The Cluster of Excellence for Fundamental Physics ``Origin and Structure of the Universe'' (Garching and Munich).

\providecommand{\bysame}{\leavevmode\hbox to3em{\hrulefill}\thinspace}

\end{document}